\documentclass{article}
\usepackage{INTERSPEECH2021}
\usepackage{amsmath}
\usepackage{graphicx}
\usepackage{tabularx}
\usepackage{multirow}

\title{End-to-End Whisper to Natural Speech Conversion using Modified Transformer Network}
\name{Abhishek Niranjan, Mukesh C. Sharma, Sai Bharath Chandra Gutha, M. Ali Basha Shaik}
\address{Samsung Research Institute, Bangalore \email{\{a.niranjan, s.chandrakan, s.gutha, m.shaik\}@samsung.com}}
\newcommand*{\addheight}[2][.5ex]{%
  \raisebox{0pt}[\dimexpr\height+(#1)\relax]{#2}%
}
\begin{document}
%\ninept
%
\maketitle
%cc
\begin{abstract}
\vspace{-2mm}
Machine recognition of an atypical speech like whispered speech, is a challenging task. We introduce whisper-to-natural-speech conversion using sequence-to-sequence approach by proposing enhanced transformer architecture, which uses both parallel and non-parallel data. We investigate different features like mel frequency cepstral coefficients and smoothed spectral features. The proposed networks are trained end-to-end using supervised approach for feature-to-feature transformation. Further, we also investigate the effectiveness of embedded auxillary decoder used after N encoder sub-layers, trained with the frame-level objective function for identifying source phoneme labels. We show results on opensource wTIMIT and CHAINS datasets by measuring word error rate using end-to-end ASR and also BLEU scores for the generated speech. Alternatively, we also propose a novel method to measure spectral shape of it by measuring formant distributions w.r.t. reference speech, as formant divergence metric. We have found whisper-to-natural converted speech formants probability distribution is similar to the groundtruth distribution. To the authors' best knowledge, this is the first time enhanced transformer has been proposed, both with and without auxiliary decoder for whisper-to-natural-speech conversion and vice versa.
\end{abstract}

\vspace{-2mm}
\section{Introduction}
%\vspace{-2mm}
Atypical speech is a kind of speech containing high level of variability mainly in temporal, spectral and prosody related patterns compared to spontaneous speech. Whispered speech, hearing impaired speech, children's speech, dysarthric speech, Autism Spectrum Disorder (ASD) Speech, Amyotrophic Lateral Sclerosis (ALS) speech, toddler's delayed speech, ageing speech etc. could be considered as various kinds of atypical speech ~\cite{aspeechintro}. Atypical Speech Recognition is one of the open and challenging issue for automatic speech recognizers (ASRs)~\cite{google2019,splissue,Splissue2014,googleproj,basics}. On the other hand, one of the key issue in atypical speech recognition include lack of sufficient amounts of training data, both in labelled and unlabelled scenarios. Thereby, voice conversion approach can be used to aid existing ASR systems, to make them more robust w.r.t. recognition accuracy~\cite{VC1,VC2}. In this work, our motivation is to improve the machine intelligibility of one type of atypical speech ie., whispered speech in the context of automatic speech recognition in E2E framework. We convert whispered speech to natural speech as part of voice conversion process and then investigate it with E2E ASR system to generate high accuracy transcriptions, with no or minimal degradation under natural speech conditions, as well.  
%due to high variability in articulation patterns.
%variability in articulation patterns

Whispered speech is a low-energy pronunciation which does not involve vocal cord vibration in it's production process and hence does not contain the fundamental frequency (F0) and also exhibit high variability in formant locations and shifts unlike spontaneous speech, thereby whispered speech exhibit noise-like characteristics~\cite{reconstuctionMorris}. 
% In general, natural speech is not appropriate form of communication in places like libraries, meeting rooms and hence people usually rely on whispered speech for human-human dialogue or human-computer interactions. Mainly, because of privacy and confidential reasons, people sometimes prefer whispered speech for communication in public places as well. 
% In addition to this, whispering is the only form of communication for patients suffering from chronic disease related to larynx structures~\cite{ito2005analysis},~\cite{Jovicic2008AcousticAO}. \\
% In principle, whispered speech is a low-energy signal as there is no vocal cord vibration involved in its production process. 
% Because the signal does not contain the fundamental frequency (F0), whispered speech signals display noise-like characteristics~\cite{reconstuctionMorris}. \\
% The energy intensity of a whisper signal is generally 20 dB lower than voiced speech signals and thus the possibility of noise interference is also higher which makes the whisper speech recognition a very challenging problem. In addition, formants are highly displaced (i.e formant shifts) compared to its corresponding natural speech signal.  \\
% Recent times have seen a surge in research on whisper speech communications. 
Whisper speech recognition have been an area of focus since the visible prominence of voice assistants like Google, Amazon's Alexa, Apple's Siri, Samsung's Bixby etc. In the literature,~\cite{xueqin2016performance} proposed the usefulness of spectrum sparse-based approach to obtain features for HMM speech recognizer model.~\cite{grozdic2017whispered} adopted deep neural networks to produce robust cepstral features to improve whispered speech recognition.
% Reference~\cite{speakerIdentificationFan}  investigated alternative feature extraction algorithm for natural/whisper speaker identification problem. Article~\cite{deng2016exploitation} exploited phrase length based features for whispered speech emotion recognition task. 
~\cite{ghaffarzadegan2016generative,whisperasrcgan,whisp20sltda} studied the inverted problem natural speech to whisper conversion for the purpose of augmenting the limited transcribed data-set for whispered speech recognition.  
%Recently, whisper-to-natural speech conversion task has gathered the attention of many researchers.
%The motive behind this research is to improve the intelligibility and ASR recognition quality of whispered speech utterances.
% \begin{figure}[!htbp]
% 	\centering
% 	\includegraphics[width=9.0cm, height=2.8cm]{flowchart}
% 	\caption{ \label{fig:transformer} Motivation behind Whisper-to-Natural speech Conversion}
% \end{figure}
Two methods were addressed in~\cite{lian2019whisper} for whispered speech conversion. Firstly, rule-based whisper conversion in which, mixed excitation linear prediction, linear prediction coding, and code excited linear prediction parameters of the source-filter model are modified on the basis of statistical differences between acoustic features of whispered and natural speech~\cite{mcloughlin2013reconstruction, yang2001linear, ahmadi2008analysis, mcloughlin2015reconstruction}.
% , sharifzadeh2009regeneration, ferreira2016implantation}. 
It is based on  simple transformation rules in combination with fundamental statistical modeling, and thereby the whisper converted speech lacks high level of intelligibility and naturalness, compared to natural speech. 
%The other methods for whisper-to-natural speech conversion includes supervised learning framework. 
On the other hand, Gaussian mixture models (GMMs)~\cite{toda2012statistical} and neural networks~\cite{li2014whisper} have been explored by researches to train a learning model using parallel training data. 
% ~\cite{toda2012statistical} built a GMM model to learn the joint spectral feature space for parallel whisper and natural speech signals. 
However,~\cite{ahangar2017voice} showed that the speech signals learned by basic GMM models display discontinuity and over-smoothing. 
%Recently, neural networks have emerged as a boon to supervised learning problems as they can learn complex nonlinear functions fairly easily. 
% \cite{li2014whisper} proposed restricted Boltzmann machine (RBM) to model joint feature space composed of whispered and parallel natural speech. 
~\cite{meenakshi2018whispered} proposed a deep bidirectional long short term memory (DBLSTM) network for speech conversion which was trained on frame-aligned parallel data and produced results which were more natural and similar to natural speech. Recently, ~\cite{lian2019whisper} proposed a sequence-to-sequence framework with fundamental network consisting of LSTM units and two separate LSTM networks to learn F0 and aperiodic component of the target natural speech, respectively. %~\cite{DBLP:journals/corr/abs-1904-06037} proposed Translatotron, an end-to-end direct speech-to-speech translation model.
\vspace{-4mm}
\section{Prior Work and Novelty}
\label{ref:novelty}
\vspace{-2mm}
In this work, we propose a novel whisper-to-natural speech conversion framework, where both parallel and non parallel data can be used for efficient training. We enhance the conventional transformer network~\cite{vaswani2017attention} and add an auxiliary decoder after N sub-layers at the encoder which is trained with the objective of identifying the tri-phone unit per frame during training. The network takes frame-level acoustic features of whispered speech as input and generates corresponding features of the target natural speech. Our model learns to map a contiguous segment of $k$ input frames to the corresponding contiguous segment of $k$ frames in output audio. Thereby an entire source audio frame sequence as input to predict the entire target audio frame sequence as output is avoided and is suitable for online whisper to natural speech conversion. %~\cite{DBLP:journals/corr/abs-1904-06037}.
Our work is distinct from ~\cite{DBLP:journals/corr/abs-1904-06037}, where, they proposed integration of two LSTM based auxiliary decoders for phoneme-sequence predictions using complete utterance at source and target sides, respectively, for offline speech translation task. We, however, incorporate only a single multi-head attention powered auxiliary decoder which predicts phoneme-units at fine resolution at frame level within enhanced transformer framework for whisper to natural-speech conversion task. As our proposed architectures uses multihead and self attention principles, they are time-efficient both in training and decoding latencies compared to existing sequential (ie., LSTM) utterance level speech-to-speech conversion systems. 
In this work, we propose two approaches for whisper-to-natural speech conversion and vice-versa. In the first method, we use MFCCs to train our end-to-end network. In our second method we use smoothed spectral features. 
%In both, the auxiliary decoder is pre-trained on opensource LibriSpeech corpus' features and corresponding tri-phone units.  
Alternatively, we also investigate natural speech to whisper conversion with our proposed architectures, as additional experiments. In principle, synthetic whisper signal can be generated by removing F0 from speech signal. However, it differs from actual whisper signal w.r.t formant shift, although perceptually, it sounds as one.
Finally, we also propose and provide  objective evidence on generated speech quality by introducing and measuring Formant Divergence Metric (FDM). As part of  indepth analysis we compute our model's ability to learn latent formant distributions using FDM, even without explicitly defining an objective function for the same (ie. formants) during training. 
\vspace{-2mm}
\section{Network Architecture}
%\label{ref:arch}
%\vspace{-4mm}
We improve the conventional transformer architecture~\cite{vaswani2017attention} to perform sequence-to-sequence modeling by omitting the obsolete embedding and positional encoding layers and adding an auxiliary decoder which takes input after first $n$ layers of encoder as shown in Fig.~\ref{fig:transformer}. Sub-components of our network are: 
\vspace{-3mm}
\subsection{Encoder}
%\vspace{-3mm}
 Encoder is composed of a stack of N = 6 identical layers.
% % like in the conventional transformer.  
 Each layer is made of two sub-layers; Multi-head self-attention layer followed by fully connected feed-forward network as shown in Fig.~\ref{fig:transformer}. Each sublayer has a residual connection and undergo layer normalization operation. All sub-layers in the model generate $Y_n$ dimensional outputs to resonate with the residual connections. For a batch size = $B$, the encoder takes input $x$ of shape $(B, k, X_n)$ and generates $z$ of same shape.    
 \vspace{-3mm}
 \subsection{Decoder}
\begin{figure}[!htbp]%[!htbp]
    %\vspace{-5mm}
	\centering
	\includegraphics[width=8cm, height=8cm]{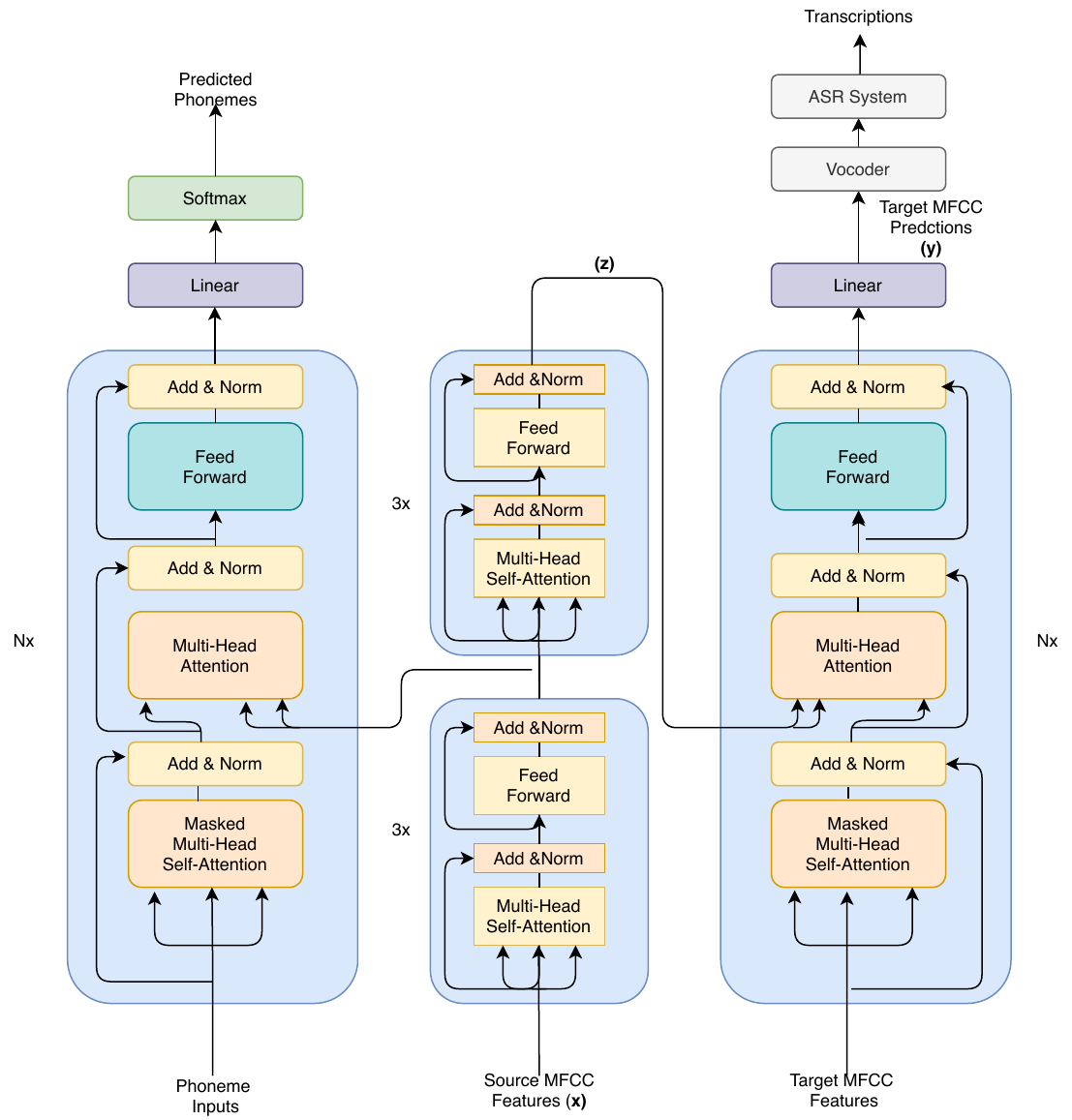}
	\caption{\label{fig:transformer} Proposed Enhanced Transformer Architecture.}
	{\vspace{-5mm}} %\vspace{-5mm}
\end{figure}

 %\vspace{-3mm}
 Akin to the encoder, decoder is also composed of N = 6 identical layers. Decoder layer is comprised of three sub-layers; Multi-head self-attention layer, encoder-decoder multi-head attention layer and feed-forward layer. Encoder-decoder sub-layer performs attention operation over the representation $\textbf{z}$ as shown in Fig.~\ref{fig:transformer}. The sub-layer connections are exactly similar to the encoder. However, we removed the softmax layer and the linear layer sits as the ultimate layer. The decoder generates $y$ of shape same as the input $x$, i.e $(B, k, Y_n)$.
% \vspace{-3mm}
 \subsection{Auxiliary Decoder}
% % ~\cite{DBLP:journals/corr/abs-1904-06037} trained two auxiliary decoders while converting speech from one language to another with an objective to predict phoneme units for both Spanish and English speech signals and reported better performance of the end-to-end model. Thus, 
% We add an auxiliary decoder to our network architecture which takes input, $\textbf{h}$, after three layers of encoder as shown in Fig.~\ref{fig:transformer} which is of shape $(B, k, X_n)$. N=3 layers in auxiliary decoder are stacked in the similar fashion as main decoder. \\
The purpose of auxiliary decoder is to predict triphone unit per source frame input.  Forced aligned triphones are generated using Kaldi (s5 recipe) ASR system~\cite{Povey_ASRU2011} for the source speech signal. On top of the librispeech lexicon, all the words contained in the training data are included in the lexicon. Similarly, training text is also included in the count based backoff $4$-gram language model.
% % We used scripts provided by Kaldi's standard s5 recipe. \\
 On the other hand, $\forall$ input $x$ of shape $(B, k, X_n)$, the tensor used to calculate cross-entropy loss over is $p$ of shape $(B, k, P)$ where $P$ is unique tri-phone vocabulary.
% \noindent We use scaled dot-product attention mechanism as described in the original transformer paper~\cite{DBLP:journals/corr/abs-1904-06037}.
We used scaled dot-product attention mechanism~\cite{DBLP:journals/corr/abs-1904-06037} shown as: 
\vspace{-2mm}
 \begin{align}
 Attention(Q,K,V) = softmax(\frac{Q.K^{T}}{\sqrt{d_k}})V 
 \end{align}
 %\vspace{-2mm}
 Where Q, K and V are query, key and value vectors respectively. $d_k$ is the dimension of key vector. We empirically selected the number of heads as 8, in the Self and Multi-head attention layers in the each of encoder, decoder and auxiliary decoder, respectively. \\ 

\vspace{-6mm}
\section{Datasets and Feature Extraction}
\label{ref:dataset}
%\vspace{-2mm}
\subsection{Transformer Model Training}
%\vspace{-3mm}
We train our network on combined dataset of SpeechOcean (King-ASR-066 American English ASR Corpus) and publicly available Whispered TIMIT (wTIMIT) corpus. 
We selected a subset of SpeechOcean data comprising of 100 hours. We professionally recorded 100hrs of American English whisper speech for the corresponding speech corpus from SpeechOcean\footnote{This whispered data is licensed and not currenlty public}, and we refer both King-ASR-066 speech corpus and recorded whisper corpus jointly as SWPC-066 (i.e. Speech-Whisper parallel corpus-066). % The data-set comprises of 84K parallel natural and whisper speech utterances having a total of 198 speakers. %were used to utter a randomly selected sentence, in both whispered and natural fashion, from a corpus of
SWPC-066 has 84K phonetically balanced sentences, wTIMIT has 450 phonetically balanced sentences of the TIMIT prompt set, resulting in 18620 parallel speech utterances (15 hours).
% Whisper speech signal and it's natural counterpart were sampled at 44kHz and 16kHz respectively, with 16-bit resolution storage. 
% , with 16-bit resolution storage. The vocabulary file generated from the transcriptions contain 23.5K unique tokens. Thereby, each utterance has on an average of 8 words and time duration of 4.27 seconds. \\
% TIMIT is a well-known corpus often used as a benchmark for phoneme recognition~\cite{timitphoneme} task. 
% The wTIMIT corpus has two accents, i.e Singaporean-English and North American, with 20 and 28 speakers from each accent group respectively. 
% The speaker recorded natural utterance and the whispered counterpart from 
% The vocabulary contains 3588 unique tokens and each utterance, on average, has 7 words. Each source-target utterance pair was sampled at 44kHz, with 16-bit resolution.\\
As our training data has audio files of different sample rates, we down-sample natural-speech audio files from 44.1kHz to 16kHz. Since the proposed architecture requires matched time-duration parallel data, we do fundamental pre-processing on source and target audio pairs. Initial and final silences are trimmed. The duration of the original audio pairs differ marginally to few mili-seconds. Time stretching operation is performed on the audio file of lower duration to make time-duration matched source-target pair. 
% As transformer network is not based on recurrent architecture, synthetically adding least amount of white noise or silence to the lower duration audios instead of time-stretching operation did not help in our experiments. 
% we compare source and target speech signals and time stretch the audio file which is of lower duration to make time-duration matching source-target pair. 
We generate 80 dimensional mel frequency cepstral coefficients (MFCC) features for the aligned dataset using publicly available Librosa toolkit~\cite{brian_mcfee_2020_3606573} and 24 dimensional smooth spectral features using WORLD Vocoder toolkit\footnote{https://github.com/JeremyCCHsu/Python-Wrapper-for-World-Vocoder}~\cite{worldvocoder}. 
The frame length was set to 25ms with 10ms overlap. As speaker information is already present in these features, we do not use any additional speaker specific features. 
As our motivation is mainly to improve the ASR accuracy under whispered speech conditions, intonation and prosody features are out of scope in this work.

\vspace{-3mm}
\subsection{Testing Model Performance} 
%\vspace{-3mm}
We used 70 parallel whisper audio files from wTIMIT test corpus and 1332 parallel whisper audio files from CHAINS corpus~\cite{chains} to test the performance of our models.
% consisting recorded audios from speakers of the two accent groups, i.e. Singaporean-English and North American, to produce MFCC features for corresponding natural speech signal.\\
% In addition, we have used CHAINS~\cite{chains} corpus as well for evaluation which contains 1332 parallel  whisper-natural audios. All audio recordings are sampled at 44.1 kHz and 16 bit PCM.
% The corpus features approximately 36 speakers (16 male, 16 female) recorded under a variety of speaking conditions. 
% In order to cover good phonetic coverage, the corpus consists of 4 fables and 33 sentences. Both fable and sentences were recorded in natural speech and whisper condition, generating 1332 parallel audios. All audio recordings are sampled at 44.1 kHz and 16 bit PCM. 
% For both test corpus we used Librosa vocoder and World vocoder to generate natural audio signals from the output MFCC and Spectral features respectively.\\
% We used following metrics to measure the quality of our generated speech signals. %To verify the intelligibility and recognition aspects of the generated audio signals we used Mean Opinion Score (MOS). 
%~\cite{zeyer2018:asr-attention}
We used word error rate to measure ASR performance on the generated natural speech and whisper signals. We also computed standard BLEU scores as translation quality metric. In addition, we measure KL divergence between reference and generated speech's formants as formant divergence metric (FDM). %distributions 
We compared the performance of ASR systems on original whisper speech signals as well as generated whispered-to-natural speech signals. 
% The ASR systems adopted and trained are described in the Section~\ref{ref:ASRresults}. 
We've used the open-source RETURNN E2E ASR~\cite{zeyer2018:asr-attention} recipe\footnote{https://github.com/rwth-i6/returnn-experiments/tree/master/2018-asr-attention/librispeech/full-setup-attention} without using the language model for fair comparison.

\vspace{-3mm}
\section{Experiments}
\label{ref:exp}

%\vspace{-3mm}
\subsection{Training Model Variations}
%\vspace{-3mm}
We train our proposed architectures with different input-output feature characteristics as shown in Table~\ref{tab:trainMod}.
% \begin{itemize}
%     \item MFCC Features: We generate 80 dimensional features for source and target speech frames using Librosa toolkit.
%     \item Smoothed Spectral Features: We generate 24 dimensional features for source and target speech frames using world vocoder toolkit. Increasing the input feature dimension didn't show noticeable improvements in the model performance. %While using these features, we add a linear layer at the start of encoder to convert the 24 dimensional vectors to 80 dimension. Similarly we append a final linear layer before computing the root mean square error loss at the end of decoder to convert back the 80 dimensional outputs to 24 dimension.
% \end{itemize}
The models $W1$ and $V1$ are our baseline transformer systems. We use root mean square loss as our objective function. The loss per input-target pair can be formulated as in Eq.\ref{eqone}, where, $y_i$ and $t_i$ represents the output vector generated for frame $i$ and the corresponding ground truth vector, number of frames($k$) is empirically set to 3 for both the models and $n$ is the dimension of output vectors, 80 for $W1$ and 24 for $V1$, respectively.
Similarly, models $W2$ and $V2$ are trained, where auxiliary decoder is also used for whisper-to-natural speech conversion. We pretrain this network using randomly sampled 30h of speech from Librispeech\footnote{We've used a subset of the corpus due to computational resource constraints}, to initialize encoder and auxiliary decoder parameters. Thereby, the use of non-parallel data further enriches the hidden representations. %\textcolor{orange}{Mukesh, write about dataset for pre-training, full libri used or part? write the number of hours used for pre-training}. 
The auxiliary decoder is trained using cross-entropy objective function for classifying tri-phone unit per frame. This loss value per input-triphone pair can be formulated as Eq.\ref{eqtwo}. %Eq.2.
%\vspace{-3mm}
\begin{align}
L_{1} &= \Big(\Sigma_{i=1}^{k}\Big)\Big(\sqrt{\frac{1}{n}\Sigma_{j=1}^{n}{({t^{j}_{i} -y^{j}_{i}})^2}}\Big) \label{eqone} \\
L_{2} &= \Big(\Sigma_{i=1}^{k}\Big)\Big(-\Sigma_{j=1}^{P}{({d^{j}_{i}log(p^{j}_{i})}})\Big) \label{eqtwo} \\
L&=L_{1}+L_{2} \label{eqthree}
\end{align}
%\vspace{-5mm}
%\begin{equation}
%\begin{split}
%L = \Big(\Sigma_{i=1}^{k}\Big)\Big(\sqrt{\frac{1}{n}\Sigma_{j=1}^{n}{({t^{j}_{i} -y^{j}_{i}})^2}}\Big) + \\ \Big(\Sigma_{i=1}^{k}\Big)\Big(-\Sigma_{j=1}^{P}{({d^{j}_{i}log(p^{j}_{i})}})\Big) 
%\end{split}
%\end{equation}

% \begin{align}
% \vspace{-4mm}
% L_{2} &= \Big(\Sigma_{i=1}^{k}\Big)\Big(-\Sigma_{j=1}^{P}{({d^{j}_{i}log(p^{j}_{i})}})\Big)\\
% L &= L_{1} + L_{2}
% \end{align}
% \vspace{-2mm}
Where, $d_i$ is one hot vector of dimension $P$ representing the tri-phone unit of frame $i$, $p_i$ is the softmax vector generated by auxiliary decoder corresponding to the frame $i$. The total loss propagated through the network is described in Eq.\ref{eqthree} for $W2$ and $V2$ models. In addition to these systems, we also train $W3$ and $V3$, which follow the same architecture as $W2$ and $V2$ respectively, for natural-speech to whisper speech conversion. 
%The use of auxiliary decoder to predict phonemes is intended to boost the enriched learning of hidden feature representations.
%In principle, our auxiliary decoder output can be further used to generate transcriptions like ASR, however, it is out of scope in this work as it requires joint training using multi-task learning approach, as one of the possibility. We mainly focus on feature to feature transformation to generate natural speech from whisper signal, as we intend to use existing ASR systems to generate transcriptions under whisper conditions, with no or minimal degradation's under natural speech conditions.

% The data-flow from the start of encoder stack to the end of auxiliary decoder stack could be thought of as DNN based ASR for whisper speech and hence the enhancement of hidden feature representations will be much more substantial than realizing a good whisper ASR accuracy~\cite{gokul_whisper}. For these reasons, we refrain from using our network as both a whisper-to-natural speech conversion system as well as a whisper ASR. 
\begin{table}[!htbp]
%\vspace{-5mm}
\begin{center}
\caption{Trained model Variations (naturalS: natural Speech)}
\label{tab:trainMod}
%\vspace{-3mm}
% \begin{tabular}{|m{1.5cm}| m{3.0cm} | m{0.95cm} | m{0.95cm} | m{1cm} | }
\begin{tabular}{|m{0.6cm}|p{1.7cm}|p{1.1cm}|p{1.1cm}|m{0.55cm}|}
\hline
Sys. & Features & Source & Target & Aux Decoder  \\ \hline
\hline
 W1 & MFCC (80) & Whisper & NaturalS & No  \\ \cline{1-1}\cline{3-5}
 W2 &  &  Whisper & NaturalS & Yes \\ \cline{1-1}\cline{3-5}
 W3 &  &  NaturalS & Whisper & Yes  \\ \hline
 V1 & Spectral    & Whisper & NaturalS & No  \\ \cline{1-1}\cline{3-5}
 V2 & Features (24)  &  Whisper & NaturalS & Yes \\ \cline{1-1}\cline{3-5}
 V3 &   &  NaturalS & Whisper & Yes  \\ %\hline
\hline
\end{tabular}
\end{center}
\vspace{-5mm}
\end{table}

%\vspace{-4mm}
\subsection{Model Parameters}
%\vspace{-3mm}
We've used the hyper-parameter settings as in~\cite{vaswani2017attention}. Adam optimizer~\cite{kingma2014adam} values are set with $\beta_{1} = 0.9$, $\beta_{2} = 0.98$ and $\epsilon = 10^{-9}$. The dropout probability ($P_{drop}$) was set to 0.1. Encoder and decoder stack consists of six (ie., N=6) identical layers each. Three (ie., N=3) identical layers are stacked to create auxiliary decoder. The number of heads used to split before applying the attention mechanism is empirically set to eight. The model variations listed in Table~\ref{tab:trainMod} were trained for $80$ epochs. We trained our models on one machine with eight NVIDIA P40 GPUs. Each training epoch took about an average of 4200 $seconds$.

% The learning rate was varied according to the formula: 
% \begin{align}
% lrate = d_{model}^{-0.5}\cdot \Gamma
% \end{align}
% where
% \begin{align}
%     \Gamma = min(step\_num^{-0.5}, step\_num\cdot warmup\_steps^{-1.5})
% \end{align}

% For the network parameters, the number of layers in encoder and decoder is set to 6. Auxiliary decoder contains a total of 3 layers. Outputs of each sub-layer are either 80 or 24 dimensional vectors corresponding to MFCC or Smoothed spectral feature inputs respectively. The number of heads used to split before applying the attention mechanism is empericaly set to 8. \\
 %\textcolor{orange}{Mukesh check this}

\vspace{-2mm}
\subsection{Results and Analysis}
\label{ref:ASRresults}
%\vspace{-3mm}
% To determine the performance of our whisper-to-natural speech conversion models, we trained multiple ASR systems~\cite{zeyer2018:asr-attention} without language model using the returnn toolkit. 
We show word error rates (WERs) on our E2E ASR systems (ie. MFCC based\footnote{As smoothed spectral features WERs are inferior compared to MFCCs, only MFCC based WERs are shown in Table~\ref{tab:Ssystems} due to page constraints.}) w.r.t. used training data tested on publicly available librispeech corpus in Table~\ref{tab:Ssystems}.
\begin{table}[!htbp]
\vspace{-2mm}
\caption{E2E ASR: Word error rate (\%) comparisons.}
\vspace{-4mm}
\centering
\label{tab:Ssystems}
\begin{tabular}{ | m{0.4cm} | m{2.6cm} | m{0.55cm} | m{0.55cm} | m{0.55cm} | m{0.55cm} | } 
\hline
ASR & Training Data & dev-clean & dev-other & test-clean & test-other\\ \hline
  \hline
S1 &  LibriSpeech (960h) & 4.87 & 14.37 & 4.87 & 15.39 \\ \hline
S2 &  LibriSpeech (960h) + wTimit + SWPC-066 (100h speech only)  & 4.48 & 14.07 & 4.63 & 14.49 \\ \hline
S3 &  LibriSpeech (960h) + wTimit + SWPC-066 (100h speech, 100h whisper)  & 4.53 & 13.90 & 4.67 & 14.40 \\ \hline
S4 &  SWPC-066 (100h speech only)  & 79 & 81.83 & 79.03 & 82.54 \\ \hline
S5 &  SWPC-066 (100h whisper only)  & 85.83 & 90.98 & 85.22 & 90.18 \\ \hline
%\multicolumn{5}{c}{As MFCCs better than smoothed spectrum features are not shown in the table}\\
% \hline
\end{tabular}
%\vspace{-4mm}
\end{table}
\begin{table*}[!htbp]
\centering
\vspace{-3mm}
\caption{E2E ASR WERs(\%) on whisper, and whisper-to-natural speech converted counterparts by Model $W1$, $W2$, $V1$ and $V2$}
\label{tab:results1}
\begin{tabular}{|c|c|c|c|c|c|c|c|c|c|c|}
\hline
                     & \multicolumn{5}{c|}{wTIMIT}                                                   & \multicolumn{5}{c|}{CHAINS}                                                   \\ \hline
\multirow{2}{*}{ASR} & \multirow{2}{*}{original whisper} & \multicolumn{4}{c|}{whisper-to-speech by} & \multirow{2}{*}{original whisper} & \multicolumn{4}{c|}{whisper-to-speech by} \\ \cline{3-6} \cline{8-11} 
                     &                                   & W1       & W2       & V1        & V2      &                                   & W1       & W2       & V1       & V2       \\ \hline
S1                   & 109.14                            & 90.86    & 80.19    & 112.38    & 92.00   & 66.35                             & 43.59    & 43.32    & 165.74   & 160.28   \\ \hline
S2                   & 106.48                            & 31.43    & 18.48    & 22.29     & 23.81   & 60.07                             & 37.64    & 37.69    & 154.54   & 150.14   \\ \hline
S3                   & 104.76                            & \textbf{22.10}    & \textbf{13.33}    & 33.71     & 34.67   & 46.29                             & \textbf{31.59}    & \textbf{32.00}    & 175.46   & 182.79   \\ \hline
\end{tabular}
\vspace{-6mm}
\end{table*}
% The ASR systems listed in Table~\ref{tab:Ssystems} were trained for 250 epochs except for S1. We picked the opensource pre-trained ASR model\footnote{https://github.com/rwth-i6/returnn-experiments/tree/master/2018-asr-attention/librispeech/full-setup-attention} described in~\cite{zeyer2018:asr-attention} which was trained for 238 epochs. 
We observe S2 and S3 systems generated lower word error rates, compared to baseline S1, on test corpora. Thus, whisper data augmentation is found to be beneficial. S4 and S5 are additional experiments showing WERs under limited training data conditions. Thus, we select S1, S2 and S3 systems for further experiments.
% The poor performance of S4 and S5 systems can be explained because of limited training data. 
% S2 and S3 systems performed better than S1 as shown in Table~\ref{tab:Ssystems}. %Besides, the SpeechOcean data has 8 words per utterance~\ref{ref:dataset} and were trained on such sequences whereas significant portion of the LibriSpeech utterances consists of $15$(Verify this number) or more words. 
As shown in Table~\ref{tab:trainMod}, we used $W1$, $W2$, $V1$ and $V2$ (ie. whisper to natural speech conversion) models to generate corresponding transformed speech features on WTIMIT and CHAINS test corpora, and use WORLD vocoder to generate its natural speech counterparts. These generated audios are tested using S1,S2 and S3 systems and WERs are shown in the Table~\ref{tab:results1}. Simplyput, Table~\ref{tab:results1} compares ASR results (using S1,S2 and S3 systems) both on original (WTIMIT/CHAINS) whisper as well as its corresponding generated speech by our enhanced transformer approaches.

\begin{figure}[!htbp]
\begin{tabular}{|c|c|}
      \hline
      \addheight{\includegraphics[width=38mm, height=40mm]{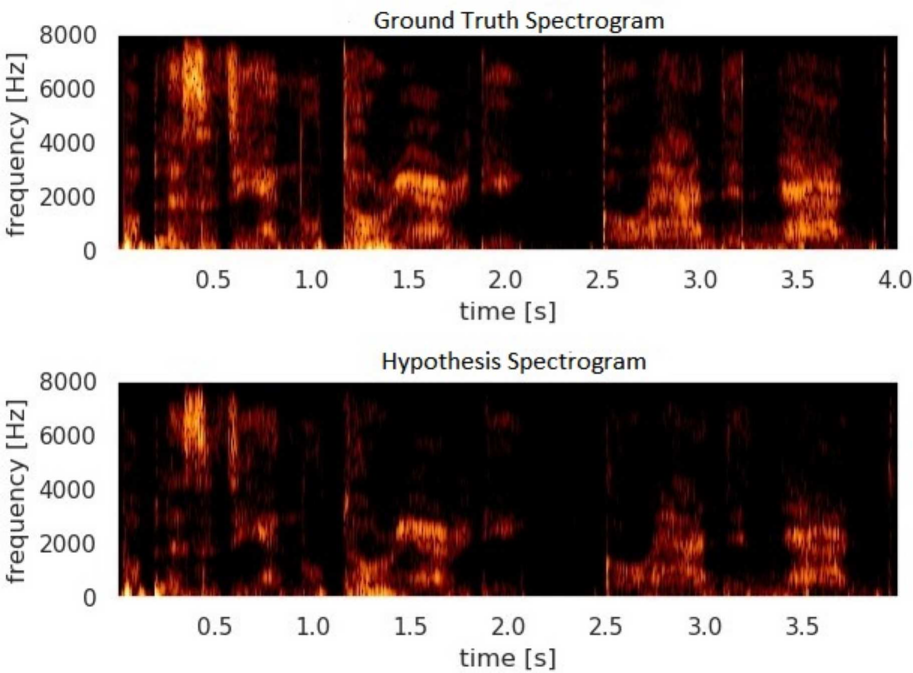}} &
      \addheight{\includegraphics[width=38mm, height=40mm]{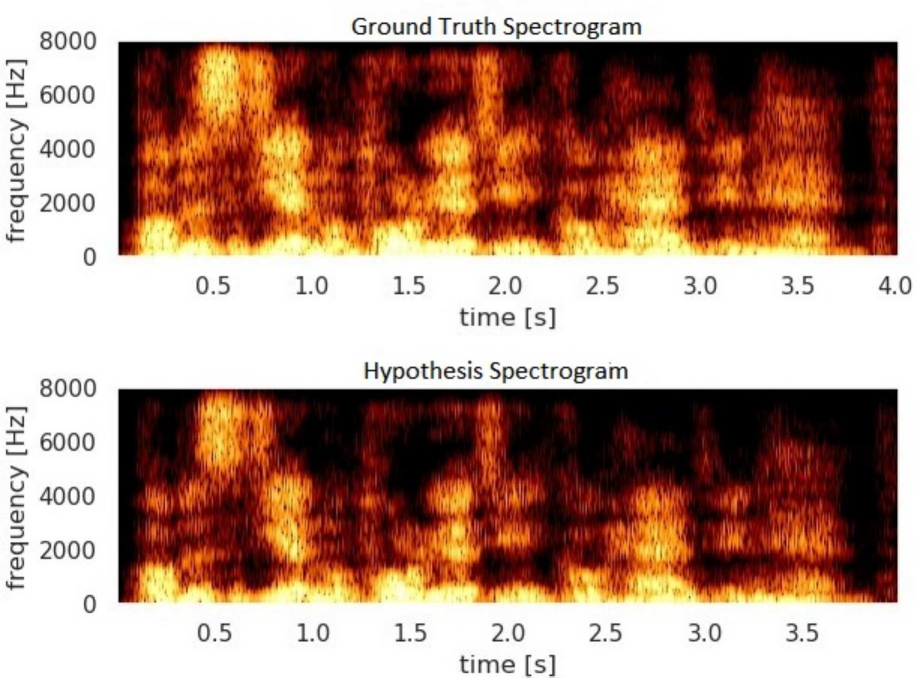}} \\
      \small Model W3 &  Model W2 \\
      \hline
      %\addheight{\includegraphics[width=38mm]{formant3_small_copy-crop}} &
      %\addheight{\includegraphics[width=38mm]{formant4_small_copy-crop}} \\
      %\small ``row 2, column 1'' &  ``row 2, column 2'' \\
      %\hline
\end{tabular}
\caption{ \label{fig:wh_w2}Spectrogram comparison (ground-truth Vs. generated). $W3$:speech-to-whisper, $W2$:whisper-to-speech}
%\vspace{-6mm}
\end{figure}

We observe ASR systems can recognize whisper-to-natural converted speech noticeably much better than the whisper utterances. We observe MFCC features performed better compared to Smoothed spectral features on both WTIMIT and CHAINS datasets, as shown in Table~\ref{tab:results1} and also in Table~\ref{tab:bleuscore}. It is also observed that, although most of the words are correctly recognized, higher number of insertions has decreased the overall accuracy of the ASR on CHAINS corpus compared to wTIMIT dataset. Whisper-to-Natural converted speech by $W2$ and $V2$ attain the lowest WERs among all the ASR systems. This can be supported with the loss inclusion from phoneme classification on source speech by auxiliary decoder during the training. We also plot spectrograms for whisper speech generated by $W3$ and output speech generated by $W2$ system along with their corresponding ground truth, respectively, as shown in Fig. ~\ref{fig:wh_w2}.
% , similarly for natural speech generated by W2 %,V2 
% system along with the corresponding ground truth natural speech depicting the model's ability in generating audio similar to the reference audios. These whispered and natural speech audios belong to wTimit dataset and are spoken by the same speaker with same transcription (``publicity and notoriety go hand in hand").

% \begin{figure}[!htbp]
% 	\centering
% 	\includegraphics[width=\columnwidth, height=8cm]{plotspect_wh_copy.eps}
% 	\caption{ \label{fig:wh_v2} Generated Whispered speech spectrogram by V3 along with the ground truth Whispered speech spectrogram}
% \end{figure}

% \begin{figure}[!htbp]
% 	\centering
% 	\includegraphics[width=\columnwidth, height=8cm]{plotspect_copy}
% 	\caption{ \label{fig:re_v3} Generated Natural speech spectrogram by V2 along with the ground truth Natural speech spectrogram}
% \end{figure}

%\begin{figure}[!htbp]
%\vspace{-2mm}
%	\centering
%	\includegraphics[width=\columnwidth, height=8cm]{plotspect_w3_copy}
%	\caption{ \label{fig:wh_w2} Generated Whispered speech spectrogram by V2 along with the ground truth Whispered speech spectrogram}
%	\vspace{-5mm}
%\end{figure}

\vspace{-7mm}
\subsection{Detailed Formant Analysis and Observations}
\label{ref:analysis}
\begin{figure}[!htbp]
\begin{tabular}{|c|c|}
      \hline
      \addheight{\includegraphics[width=35mm]{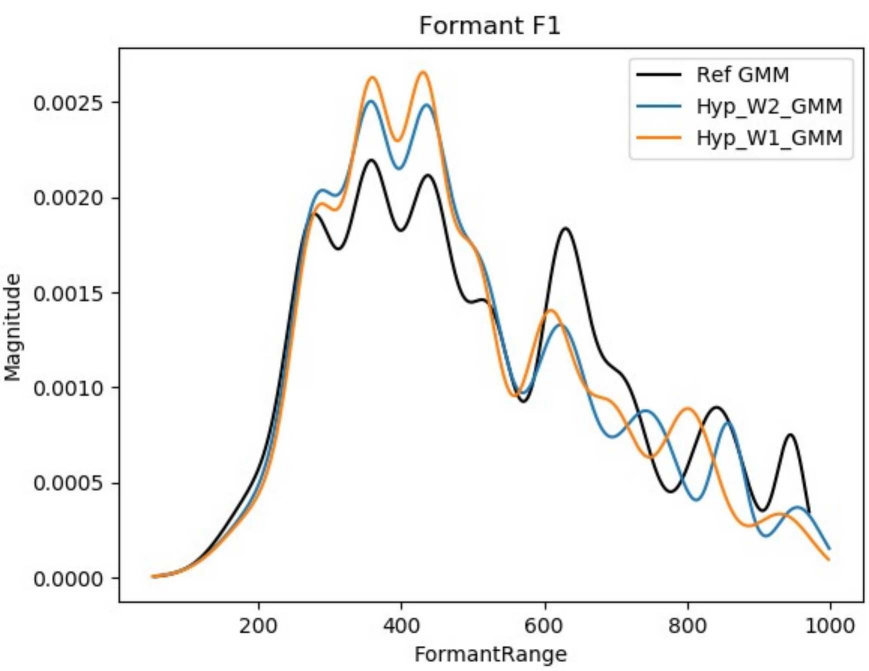}} &
      \addheight{\includegraphics[width=35mm]{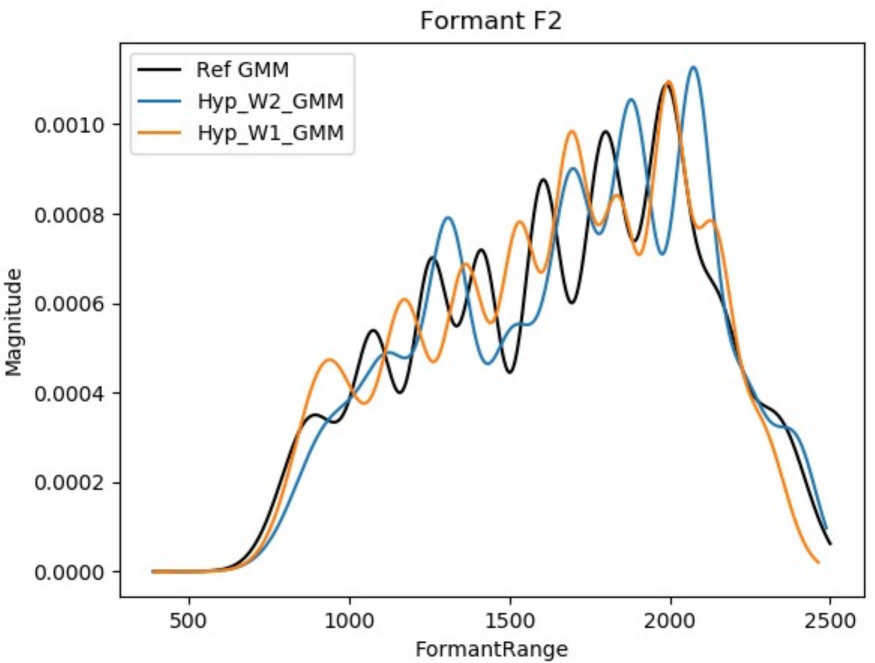}} \\
      \small Formant F1 &  Formant F2 \\
      \hline
      \addheight{\includegraphics[width=35mm]{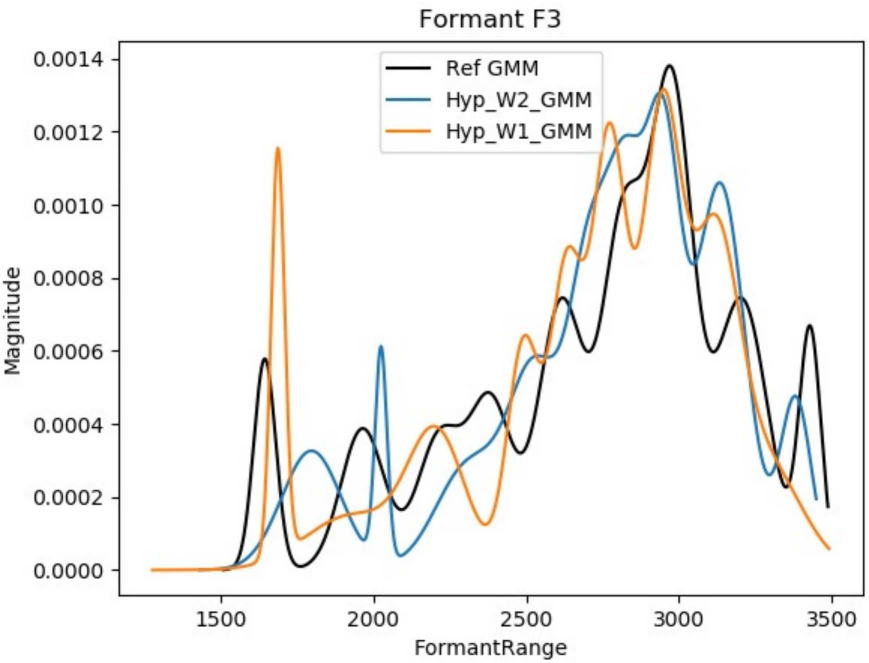}} &
      \addheight{\includegraphics[width=35mm]{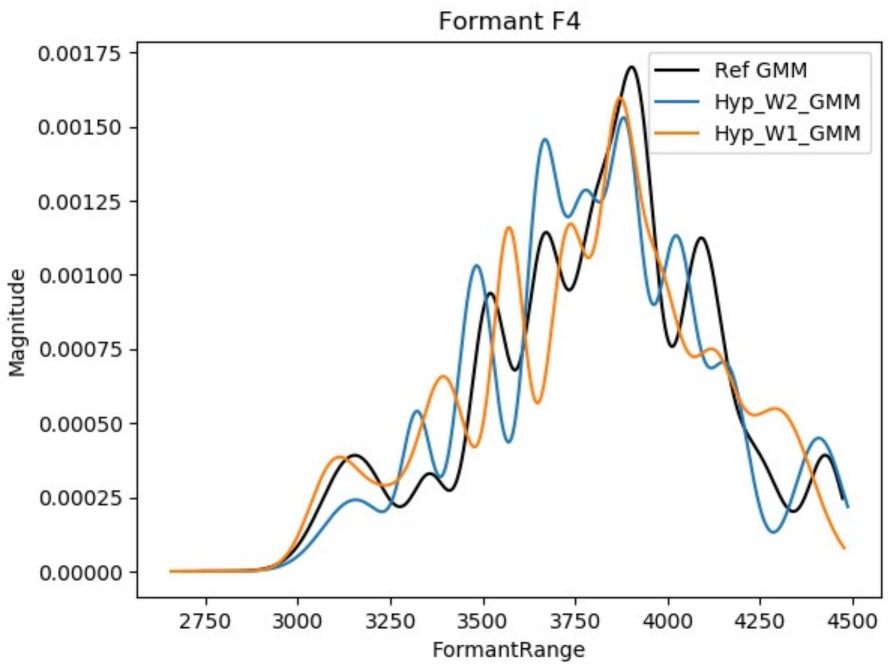}} \\
      \small Formant F3 &  Formant F4 \\
      \hline
\end{tabular}
\caption{ \label{formants} F1-F4 GMM comparisons (W1, W2 Vs. groundtruth)}
\vspace{-4mm}
\end{figure}

% \texttt{Mukesh's part here\\}
%\vspace{-3mm}
% The vocal cords are basic elements of speech production in humans. They generally produce well-defined pitch under naturally spoken conditions. A formant is the spectral shaping resulting from an acoustic resonance caused by the geometry of the physiological tubular system of the speaker's vocal tract. As we speak, the cavities of our vocal tract changes in the shape and volume, so the formant frequencies will be constantly changing. 
% The effects of formants are visible in spectrogram of a speech, because the spectrum is affected by the resonance of vocal tract. 
% In this section we study first four formant frequencies (F1, F2, F3, F4). %and fundamental frequency (F0).
%Fundamental frequency or F0 is the frequency at which vocal chords vibrate in voiced sound. F0 measurements were made because F0 is known to affect formant measurements~\cite{fundamentalf0}.
% The frequency location of F1 and F2 are primarily based on the shape of the vocal tract as tongue, jaws, lips move to generate sound. The frequency of third formant F3 is related to few specific speech sounds. The fourth (F4) and higher formants remains almost constant in frequency location regardless of changes in articulation. 
As the formant frequency locations depends on three factors, like pharyngeal-oral tract length, the location of constriction in the tract and the degree of narrowness of the constriction, we consider only dominant formants F1-F4.
%\begin{figure}[!htbp]
%\vspace{-2mm}
%	\centering
%	\includegraphics[width=\columnwidth,height=8cm]{plotspect_w2_copy}
%	\caption{ \label{fig:re_w3} Generated Natural speech spectrogram by W2 along with the ground truth Natural speech spectrogram}
%	\vspace{-5mm}
%\end{figure}
\begin{table}[!htbp]
\vspace{-3mm}
\caption{BLEU score on S3 system transcriptions for best models $W2$ and $V2$ (ie. whisper to natural speech)}
\centering
\label{tab:bleuscore}
\vspace{-3mm}
\begin{tabular}{|l|c|c|c|} 
\hline
Model & wTIMIT  &  CHAINS \\ 
  \hline
W2 & 85.36 & 62.68 \\ \hline
V2 & 68.60 & 33.00 \\ 
\hline
\end{tabular}
\vspace{-4mm}
\end{table}
We used standard Burg (1967) linear prediction coefficients (LPC) algorithm for formant frequency estimation ~\cite{burg}. In order to examine the audios generated by our model, we have analysed the behavior of %fundamental frequency (F0) 
and first four formants (F1-F4) and compared with the reference audio formants. As shown in the formants graph (i.e Fig~\ref{formants}) we fit a $k$-component Gaussian Mixture Model (GMM) on the reference audio formants (i.e. Ref GMM), where, $k$ is a hyperparameter determined by EM algorithm. We try to fit a GMM model on the formants value extracted from audios generated by our model W1 and W2 (i.e. Hyp\_GMM\_W1 and Hyp\_GMM\_W2). 
% As we can observe in the Fig~\ref{formants} that GMM fit on the formants of the generated audios has a marginal shift from the Reference GMM but follows the same pattern. 
\begin{table}[!h]
%\vspace{-3mm}
\caption{Formant Divergence Metric (FDM) using KL-Divergence (Ref. GMM Vs. (W1 or W2) GMM).}
\centering
\label{tab:kldivnorm}
\begin{tabular}{| m{1.0cm} | m{1.0cm} | m{1.0cm} | m{1.0cm} | m{1.0cm} |}
\hline
%\hline
   & F1 & F2 & F3 & F4  \\ \hline
 % \hline
 W1  & 0.1543 & 0.0260 & 0.3868 & 0.0675 \\ \hline
 W2  & 0.1034 & 0.0234 & 0.1945 & 0.0474 \\ 
\hline
\end{tabular}
\vspace{-5mm}
\end{table}
As shown in Fig~\ref{formants}, GMM of model $W2$ is more similar to the reference GMM as compared model $W1$'s GMM. Thereby, We computed FDM using KL-Divergence, as shown in Table~\ref{tab:kldivnorm},
% , tab:kldivvocodernorm} 
between Ref\_GMM and Hyp\_GMM\_W1, Hyp\_GMM\_W2 using Monte Carlo method. Simplyput, the lower the FDM, the lesser the divergence between the groundtruth and the generated speech w.r.t. the target formant distribution.
%~\cite{kldivergencegmm}. 
% KL divergence values are shown in Table~\ref{tab:kldivnorm} for comparison W1 and W2 systems. Corresponding formant graphs are shown in Fig~\ref{formants}. Lower values towards zero are always preferred. 
% We observe KL divergence values for all formants F1-F4 for W2 is better than W1, %except F0.

% Similarly, KL divergence values are shown in Table~\ref{tab:kldivvocodernorm} for comparison of V1 and V2 systems. %Corresponding formant graphs are shown in Fig~\ref{formantsvocoder}. 
As shown in the Table \ref{tab:kldivnorm}, FDM for formants F1-F4 are better for W2 and V2 than W1 and V1, respectively. Thereby, W2/V2 systems outperformed baseline systems W1/V1, respectively. The formants generated by the W2's output speech are more precise than W1 compared to the target speech formants. % On the other hand, we also observe KL divergence values generated using W3 are better than V3 system (ie. Natural-speech to whisper, \textit{cf}. Table~\ref{tab:kldivn2wnorm} ).
\vspace{-4mm}
\section{Conclusion}
\label{ref:con}
\vspace{-2mm}
In this paper, we made an attempt to improve machine intelligibility of one type of atypical speech, ie., whispered speech in the context of speech recognition. We've proposed enhanced transformer framework for whispered-to-natural speech conversion, and vice versa. We investigated the effectiveness of embedded auxillary decoder as an enhancement in transformer architecture. Our proposed approach uses both parallel and non parallel data to address data sparcity issue in atypical speech recognition. We passed whisper-to-natural converted speech through end-to-end ASR systems and validated the performance of our approaches, and also show that converted natural speech signals are noticeably more recognizable and intelligible compare to original whisper speech audios.
%Demo files are herewith attached.
% trained on different datasets. 
% Experimental results show that natural speech generated from whisper speech using our proposed architecture is more recognizable and intelligible compare to original whisper speech. 
% Our proposed models are capable of generating natural speech without taking explicit speaker dependent information as input. As our motivation is mainly to improve the ASR accuracy under whisper conditions, output speech quality is of higher importance compared to speaker related information. 
% Word error rate (WERs) drop by $\approx$ $65\%$ when a whisper utterance is converted to natural speech signal by our best sequence-to-sequence system which adds an auxiliary decoder to the fundamental encoder-decoder stacks. MFCC features performs better than smoothed spectrum features in terms of over all performance. We also showed that the synthesised natural speech from whisper signals follows the same pattern of formant distribution as the original natural speech counterpart. We plan to make our source code, used features, and pytorch model checkpoints publicly available in the near future.
Word error rates drop by $\approx$ $65\%$ when  whisper utterances are converted to natural speech signals and tested by the highly optimized E2E RETURNN ASR. We also showed that our generated natural speech from whisper signals follows the near similar pattern of formant distributions with  minimal divergence compared to the natural speech counterpart. We plan to make our PyTorch source code publicly available in the future.

\bibliographystyle{IEEEbib}
\bibliography{refs}

\begin{thebibliography}{10}

\bibitem{aspeechintro}
Georg Stemmer, Elmar N{\"o}th, and Vijay Parsa,
\newblock ``{Atypical Speech},''
\newblock in {\em EURASIP Journal on Audio Speech and Music Processing}, march
  2010.

\bibitem{google2019}
Joel Shor, Dotan Emanuel, Oran Lang, Omry Tuval, Michael Brenner, Julie
  Cattiau, Avinatan Hassidim, and Yossi Matias,
\newblock ``{Personalizing ASR for Dysarthric and Accented Speech with Limited
  Data},''
\newblock in {\em Interspeech}, Graz, Austria, 2019.

\bibitem{splissue}
Bj{\"o}rn Schuller, Tiago~H. Falk, Vijay Parsa, and Elmar N{\"o}th,
\newblock ``Atypical speech and voices: Corpora, classification, coaching and
  conversion,''
\newblock in {\em Special Issue: EURASIP Journal on Audio, Speech and Music
  Processing}. 2015, SpringerOpen.

\bibitem{Splissue2014}
Marek Bohac, Michaela Kucharova, Zoraida Callejas, Jan Nouza, and Petr
  \v{C}erva,
\newblock ``A cross-lingual adaptation approach for rapid development of speech
  recognizers for learning disabled users,''
\newblock in {\em Special Issue: EURASIP Journal on Audio, Speech and Music
  Processing}. 2014, number~39, SpringerOpen.

\bibitem{googleproj}
Google Inc.,
\newblock ``{Project Euphonia: A Research initiative focused on helping people
  with atypical speech, https://sites.research.google/euphonia/about},''
\newblock .

\bibitem{basics}
Tartter Vivien,
\newblock ``{What's in a whisper?},''
\newblock in {\em Journal of Acoustic Society of America}, 1989, vol. 86(5).

\bibitem{VC1}
Yuki Takashima, Tetsuya Takiguchi, and Yasuo Ariki,
\newblock ``{End-to-End Dysarthric Speech Recognition Using Multiple
  Databases},''
\newblock in {\em {ICASSP} 2019, Brighton, United Kingdom, May 12-17, 2019}.
  2019, pp. 6395--6399, {IEEE}.

\bibitem{VC2}
Ryo Aihara, Tetsuya Takiguchi, and Yasuo Ariki,
\newblock ``{Phoneme-Discriminative Features for Dysarthric Speech
  Conversion},''
\newblock in {\em Interspeech 2017}, Stockholm, Sweden, August 2017, pp.
  3374--3378.

\bibitem{reconstuctionMorris}
Robert Morris and Mark Clements,
\newblock ``Reconstruction of speech from whispers,''
\newblock {\em Medical engineering \& physics}, vol. 24, pp. 515--20, 09 2002.

\bibitem{xueqin2016performance}
Chen Xueqin, Zhao Heming, and Fan Xiaohe,
\newblock ``Performance analysis of mandarin whispered speech recognition based
  on normal speech training model,''
\newblock in {\em 2016 Sixth International Conference on Information Science
  and Technology (ICIST)}. IEEE, 2016, pp. 548--551.

\bibitem{grozdic2017whispered}
DJordje~T Grozdi{\'c} and Slobodan~T Jovi{\v{c}}i{\'c},
\newblock ``Whispered speech recognition using deep denoising autoencoder and
  inverse filtering,''
\newblock {\em IEEE/ACM Transactions on Audio, Speech, and Language
  Processing}, vol. 25, no. 12, pp. 2313--2322, 2017.

\bibitem{ghaffarzadegan2016generative}
Shabnam Ghaffarzadegan, Hynek Bo{\v{r}}il, and John~HL Hansen,
\newblock ``Generative modeling of pseudo-whisper for robust whispered speech
  recognition,''
\newblock {\em IEEE/ACM Transactions on Audio, Speech, and Language
  Processing}, vol. 24, no. 10, pp. 1705--1720, 2016.

\bibitem{whisperasrcgan}
Prithvi Raj~Reddy Gudepu, Gowtham~Prudhvi Vadisetti, Abhishek Niranjan, Kinnera
  Saranu, Raghava Sarma, M.~Ali~Basha Shaik, and Periyasamy Paramasivam,
\newblock ``Whisper augmented end-to-end/hybrid speech recognition system -
  cyclegan approach,''
\newblock in {\em INTERSPEECH}, 2020.

\bibitem{whisp20sltda}
Heng-Jui Chang, Alexander~H. Liu, Hung yi~Lee, and Lin-Shan Lee,
\newblock ``{End-to-End Whispered Speech Recognition with Frequency-weighted
  Approaches and Pseudo Whisper Pre-training},''
\newblock in {\em IEEE SLT}, January 2021.

\bibitem{lian2019whisper}
Hailun Lian, Yuting Hu, Weiwei Yu, Jian Zhou, and Wenming Zheng,
\newblock ``Whisper to normal speech conversion using sequence-to-sequence
  mapping model with auditory attention,''
\newblock {\em IEEE Access}, vol. 7, pp. 130495--130504, 2019.

\bibitem{mcloughlin2013reconstruction}
Ian~Vince McLoughlin, Jingjie Li, and Yan Song,
\newblock ``Reconstruction of continuous voiced speech from whispers.,''
\newblock in {\em INTERSPEECH}, 2013, pp. 1022--1026.

\bibitem{yang2001linear}
Ming Yang, Fenghai Qiu, and Fuyuan Mo,
\newblock ``A linear prediction algorithm in low bit rate speech coding
  improved by multi-band excitation model,''
\newblock {\em ACTA ACUSTICA-PEKING-}, vol. 26, no. 4, pp. 329--334, 2001.

\bibitem{ahmadi2008analysis}
Farzaneh Ahmadi, Ian~Vince McLoughlin, and Hamid~Reza Sharifzadeh,
\newblock ``Analysis-by-synthesis method for whisper-speech reconstruction,''
\newblock in {\em APCCAS 2008-2008 IEEE Asia Pacific Conference on Circuits and
  Systems}. IEEE, 2008, pp. 1280--1283.

\bibitem{mcloughlin2015reconstruction}
Ian~V Mcloughlin, Hamid~Reza Sharifzadeh, Su~Lim Tan, Jingjie Li, and Yan Song,
\newblock ``Reconstruction of phonated speech from whispers using
  formant-derived plausible pitch modulation,''
\newblock {\em ACM Transactions on Accessible Computing (TACCESS)}, vol. 6, no.
  4, pp. 1--21, 2015.

\bibitem{toda2012statistical}
Tomoki Toda, Mikihiro Nakagiri, and Kiyohiro Shikano,
\newblock ``Statistical voice conversion techniques for body-conducted unvoiced
  speech enhancement,''
\newblock {\em IEEE Transactions on Audio, Speech, and Language Processing},
  vol. 20, no. 9, pp. 2505--2517, 2012.

\bibitem{li2014whisper}
Jing-jie Li, Ian~V McLoughlin, Li-Rong Dai, and Zhen-hua Ling,
\newblock ``Whisper-to-speech conversion using restricted boltzmann machine
  arrays,''
\newblock {\em Electronics Letters}, vol. 50, no. 24, pp. 1781--1782, 2014.

\bibitem{ahangar2017voice}
Mohsen Ahangar, Mostafa Ghorbandoost, Sudhendu Sharma, and Mark~JT Smith,
\newblock ``Voice conversion based on a mixture density network,''
\newblock in {\em 2017 IEEE Workshop on Applications of Signal Processing to
  Audio and Acoustics (WASPAA)}. IEEE, 2017, pp. 329--333.

\bibitem{meenakshi2018whispered}
G~Nisha Meenakshi and Prasanta~Kumar Ghosh,
\newblock ``Whispered speech to neutral speech conversion using bidirectional
  lstms.,''
\newblock in {\em Interspeech}, 2018, pp. 491--495.

\bibitem{vaswani2017attention}
Ashish Vaswani, Noam Shazeer, Niki Parmar, Jakob Uszkoreit, Llion Jones,
  Aidan~N Gomez, {\L}ukasz Kaiser, and Illia Polosukhin,
\newblock ``Attention is all you need,''
\newblock in {\em Advances in neural information processing systems}, 2017, pp.
  5998--6008.

\bibitem{DBLP:journals/corr/abs-1904-06037}
Ye~Jia, Ron~J. Weiss, Fadi Biadsy, Wolfgang Macherey, Melvin Johnson, Zhifeng
  Chen, and Yonghui Wu,
\newblock ``Direct speech-to-speech translation with a sequence-to-sequence
  model,''
\newblock {\em CoRR}, vol. abs/1904.06037, 2019.

\bibitem{Povey_ASRU2011}
Daniel Povey, Arnab Ghoshal, Gilles Boulianne, Lukas Burget, Ondrej Glembek,
  Nagendra Goel, Mirko Hannemann, Petr Motlicek, Yanmin Qian, Petr Schwarz, Jan
  Silovsky, Georg Stemmer, and Karel Vesely,
\newblock ``The kaldi speech recognition toolkit,''
\newblock in {\em IEEE 2011 Workshop on Automatic Speech Recognition and
  Understanding}. Dec. 2011, IEEE Signal Processing Society,
\newblock IEEE Catalog No.: CFP11SRW-USB.

\bibitem{brian_mcfee_2020_3606573}
Brian~McFee et. al.,
\newblock ``librosa/librosa: 0.7.2,'' Jan. 2020.

\bibitem{worldvocoder}
Masanori Morise, Fumiya YOKOMORI, and Kenji Ozawa,
\newblock ``World: A vocoder-based high-quality speech synthesis system for
  real-time applications,''
\newblock {\em IEICE Transactions on Information and Systems}, vol. E99.D, pp.
  1877--1884, 07 2016.

\bibitem{chains}
F.~Cummins, Marco Grimaldi, T.~Leonard, and Juraj Simko,
\newblock ``The \uppercase{CHAINS} corpus: Characterizing individual
  speakers,''
\newblock {\em Proc. SPECOM}, pp. 431--435, 01 2006.

\bibitem{zeyer2018:asr-attention}
Albert Zeyer, Kazuki Irie, Ralf Schlüter, and Hermann Ney,
\newblock ``Improved training of end-to-end attention models for speech
  recognition,''
\newblock in {\em Interspeech}, Hyderabad, India, Sept. 2018.

\bibitem{kingma2014adam}
Diederik~P Kingma and Jimmy Ba,
\newblock ``Adam: A method for stochastic optimization,''
\newblock {\em arXiv preprint arXiv:1412.6980}, 2014.

\bibitem{burg}
N.~Anderson,
\newblock ``On the calculation of filter coefficients for maximum entropy
  spectral analysis,''
\newblock in {\em Childers` Modern Spectrum Analysis}. 1978, pp. 431--435, IEEE
  Press.

\end{thebibliography}

\end{document}